\documentclass[aps,prl,twocolumn,showpacs,superscriptaddress,amsmath,amssymb]{revtex4-1}
\usepackage[utf8]{inputenc}
\usepackage{graphicx}% Include figure files
\usepackage{hyperref}
\DeclareMathOperator{\sech}{sech}

\begin{document}

%\preprint{}

\title{Microscopic nanomechanical dissipation in gallium arsenide resonators} %Title of paper



\author{M. Hamoumi}
\author{P. E. Allain}
\author{W. Hease}
%\thanks{}
\affiliation{Matériaux et Phénomènes Quantiques, Université Paris Diderot, CNRS UMR 7162, Sorbonne Paris Cité, 75013 Paris, France}

\author{L. Morgenroth}
\affiliation{Institut d’Electronique, de Microélectronique et de Nanotechnologie, UMR CNRS 8520, Avenue Poincaré, 59652, Villeneuve d’Ascq, France}

\author{B. Gérard}
\affiliation{III-V Lab, 1 Avenue Augustin Fresnel, 91767 Palaiseau, France}

\author{A. Lemaître}
\affiliation{Centre de Nanosciences et de Nanotechnologies, CNRS, Université Paris Sud, Université Paris-Saclay, C2N-Marcoussis, Route de Nozay, 91460 Marcoussis, France}

\author{G. Leo}
\author{I. Favero}
\affiliation{Matériaux et Phénomènes Quantiques, Université Paris Diderot, CNRS UMR 7162, Sorbonne Paris Cité, 75013 Paris, France}


\date{\today}

\begin{abstract}
We report on a systematic study of nanomechanical dissipation in high-frequency ($\approx 300$ MHz) gallium arsenide optomechanical disk resonators, in conditions where clamping and fluidic losses are negligible. Phonon-phonon interactions are shown to contribute with a loss background fading away at cryogenic temperatures ($3$ K). Atomic layer deposition of alumina at the surface modifies the quality factor of resonators, pointing towards the importance of surface dissipation. The temperature evolution is accurately fitted by two-level systems models, showing that nanomechanical dissipation in gallium arsenide resonators directly connects to their microscopic properties. Two-level systems, notably at surfaces, appear to rule the damping and fluctuations of such high-quality crystalline nanomechanical devices, at all temperatures from 3 to 300K.
\end{abstract}


\maketitle %\maketitle must follow title, authors, abstract and \pacs


The physical origin of nanomechanical dissipation is a topic of curiosity and debate, motivated by a vast number of applications. Ultra-low dissipation nanomechanical resonators represent a key ingredient for optomechanics, which investigates the interaction of light and mechanical motion \cite{favero2009optomechanics,aspelmeyer2014cavity}. They are becoming crucial in weak-force resolution \cite{miao2012microelectromechanically,krause2012high}, mass sensing \cite{tamayo2013biosensors,supg-toroidOE,gil2015naturenano, gil2016nano}, or mesoscopic quantum operations such as ground-state cooling of mechanical motion \cite{teufel2011sideband, peterson2016laser} and entanglement between mechanical systems \cite{borkje2011proposal}. For example Gallium Arsenide (GaAs) nano-optomechanical disk resonators, whose high-frequency radial breathing modes (RBMs) strongly couple to optical whispering gallery modes (WGMs) \cite{ding2014gallium, favero2014chapter}, are expected to display low mechanical dissipation thanks to their constitutive crystalline epitaxial material, and they indeed achieved large Q-frequency products. However, despite the achieved control of clamping losses \cite{nguyen2013ultrahigh,nguyen2015improved}, their ultimate mechanical performances are still affected by residual damping processes. The investigation of these processes is the focus of the present work. 

In this Letter, specific dissipation channels are made negligible by experimental conditions (vacuum operation that suppresses fluidic damping) or by design (pedestal engineering that suppresses anchoring losses \cite{nguyen2013ultrahigh,nguyen2015improved}), enabling a direct analysis of intrinsic loss mechanisms. These are investigated by comparative measurement of identical resonators made out of two distinct epitaxial wafers, and accurately compared to models of phonon-phonon damping. Surface nanomechanical dissipation is investigated by observing the influence of an atomic layer deposition (ALD) of alumina Al$_2$O$_3$ onto the resonators. The temperature dependence between 3 and 300 K is systematically measured and fitted by two-level systems (TLS) models, allowing the emergence of a microscopic picture of damping processes in GaAs resonators. Our results indicate that TLS dissipation dominate at all temperatures, despite the crystalline nature of the material. By comparing distinct wafers, as well as pristine and surface-treated resonators, we provide evidences about the nature and localization of TLS. Our study finally provides a consistent picture of noise mechanisms affecting high-Q crystalline nanomechanical systems, which are generally regarded as best candidates for quantum applications.
 
\begin{figure*}[htpb]
    \includegraphics[width=13.3cm]{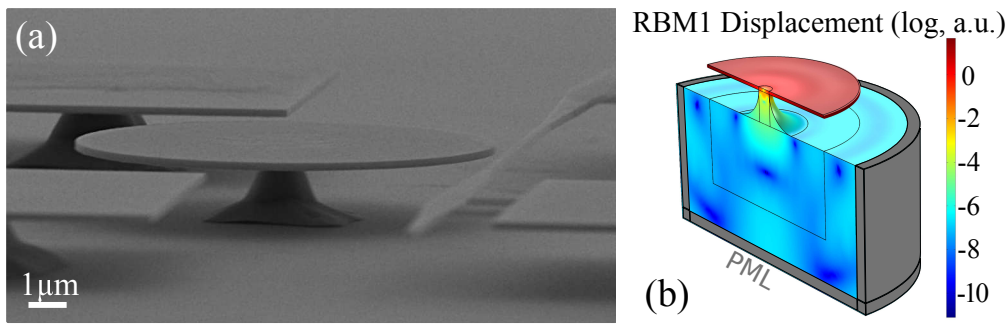}
    \caption{(a) A 200 nm thick GaAs disk resting on a 500 nm radius AlGaAs pedestal. The suspended optical coupling waveguide with inverted tapers can be seen on the right side. The square-shaped pads on the link hold the guide of the adjacent device (not shown). (b) FEM simulation of the first-order RBM of the disk. The color scale corresponds to the modulus of the displacement vector, in logarithm scale and in arbitrary units. The black lines correspond to the different geometrical domains at rest for the FEM simulation. Perfectly Matched Layers (PML) are introduced at the substrate boundaries. The displacement is strongly localized within the disk, corresponding to a clamping $\Gamma_m<2.5$ kHz ($Q_m >6.5\cdot10^5$).\label{diskfabrication}}
\end{figure*}

The employed GaAs optomechanical disks have a radius of $5.5 \ \mu$m, are $200$ nm thick, and sit on a $1.8 \ \mu$m high aluminium gallium arsenide (AlGaAs) pedestal of radius $500$ nm. They are measured by optical means. Light is brought to a disk WGM via evanescent coupling to an on-chip suspended waveguide \cite{baker2011critical}, whose endings are shaped into inverted tapers to optimize coupling to micro-lensed fibers. The samples are fabricated from two distinct wafers (1 and 2) grown by molecular beam epitaxy (MBE) under distinct conditions but with the same nominal structure: $200$ nm(GaAs)/$1.8 \ \mu$m(Al$_{0.8}$Ga$_{0.2}$As)/$500$  $\mu$m(semi-insulating GaAs). The disk and waveguides are first patterned in a negative resist using electron beam lithography. The resist is developed and serves as a mask during the Inductively Coupled Plasma Reactive Ion Etching (ICP-RIE) with a SiCl$_4$/Ar chemistry. The pedestal is under-etched with a hydrofluoric acid solution, and the tips of the inverted tapers are freed using a BCK solution \cite{TheseWilliam}. Fig. \ref{diskfabrication}a shows an electron micrograph of a fabricated device.

In the following, we invariably measure the quality factor $Q_m$ of a mechanical mode or its energy dissipation rate $\Gamma_m = \omega_{m}/Q_m$, obtained from the full line width of the corresponding resonance. The mechanical spectrum is measured optomechanically by tuning the laser on the flank of an optical resonance and analyzing the radio-frequency noise of the output light with a fast photo-detector connected to an electronic spectrum analyser \cite{ding2014gallium}. To avoid dynamical optomechanical back-action modifying $\Gamma_m$ \cite{favero2009optomechanics,aspelmeyer2014cavity}, the measurements are taken as function of optical power and the linear evolution extrapolated at zero power. We focus here on the first order RBM of the above-discussed disks, which has a frequency of $f_m= \omega_{m}/2\pi= 260$ MHz and is in our case only subject to intrinsic dissipation channels. Indeed, in this work GaAs disk resonators are operated in a cryostat (accessible range $2.6$ to $300$ K) and under vacuum ($\leq 10^{-5}$ mbar). At such pressure, the gas damping of the breathing motion is negligible \cite{verbridge2008megahertz,gil2015naturenano}. The dimensions of the disk and pedestal are also  chosen to render clamping losses negligible. The latter are simulated numerically by Finite Element Method (FEM), as shown in Fig. \ref{diskfabrication}b, and our previous work in the clamping-limited regime showed good agreement with experiments \cite{nguyen2013ultrahigh, nguyen2015improved}. We adopt here a disk for which our tolerance on pedestal dimensions bounds clamping losses $\Gamma_m$ to below $3.26$ kHz, corresponding to a $Q_m > 5\cdot10^5$. In what follows, this channel of dissipation can be neglected, whatever the temperature.

\begin{figure}[htbp]
%  \begin{minipage}[c]{13.3cm}
    \includegraphics[width=6.4cm]{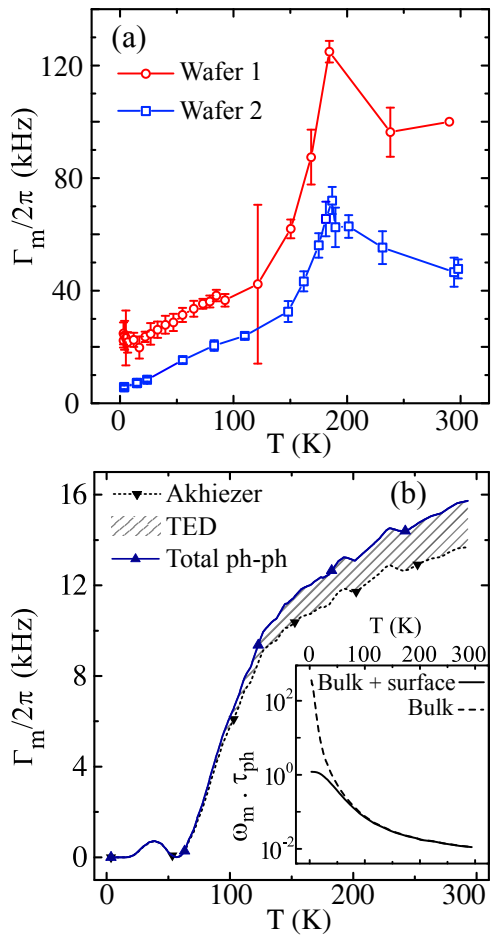}
%  \end{minipage}\hfill
%  \begin{minipage}[c]{0.25\textwidth}
	\caption{(a) Measured mechanical dissipation rate as a function of temperature. The two wafers were grown in two distinct MBE chambers. (b) Modeled mechanical dissipation due to phonon-phonon interactions, for both Akhiezer and TED mechanisms. Note that the damping cancels artificially around 60K with the Grüneisen parameter, as a result of our approximation on $\Delta\gamma$ (see text). Inset: Temperature evolution of the parameter $ \omega_{m}\tau_{ph}$ (see text).
\label{measurementsTED}}
%  \end{minipage}
\end{figure}

The measurements of $\Gamma_m$ between $3$ and $300$K are shown in Fig. \ref{measurementsTED}a for nominally identical resonators fabricated with the exact same process, but out of the two distinct epitaxial wafers (1 and 2). These results reveal two obvious features. Firstly, the intrinsic dissipation tends to increase with temperature, in a similar manner for the two wafers; secondly, the dissipation is larger in wafer 1 than in wafer 2. The temperature evolution of $\Gamma_m$ distinguishes three regimes: (1) Slow increase between $3$ K and $150$ K (2) Peak around $180$ K (3) Quasi-plateau from $200$ K to $300$ K. The similar behavior of wafers 1 and 2 points towards some universality, whose origin remains to be elucidated. Fluidic and clamping losses being negligible, the dissipation processes must take place in the bulk or at the surface of resonators.

We first analyze the mechanical dissipation induced by interaction of the $260$ MHz (mechanical) phonon with high-frequency ($\simeq h/k_{B}T$) thermal phonons. This phonon-phonon damping was discussed in the bulk, using the Landau-Rumer approach (valid when $ \omega_{m}\tau_{ph}\gg 1$) or a Boltzmann equation approach like employed by Akhiezer (valid when $ \omega_{m}\tau_{ph}\ll 1$), where $\tau_{ph}$ is a relaxation time for thermal phonons. The relation between these approaches was discussed by Maris \cite{maris1971physical}. The Boltzmann equation description assumes thermal phonons to be localized with respect to the mechanical wavelength, a condition implying that $k_B T \gg hf_m$, which is satisfied here where $T>3K$. Upon incidence of a mechanical wave (even spatially uniform), the population of thermal phonons is perturbed as consequence of the lattice anharmonicity and dissipates energy via collisions to return to equilibrium. A collision time approximation can be adopted, provided $\omega_{m}\tau_{ph}< 1$, leading to an expression of this Akhiezer damping \cite{Maris1968,maris1971physical,Perrin1981,ClelandBook}:

%The main contribution to the dissipation in this regime is that of longitudinal mechanical phonons, and the dissipation rate is shown to be:
%\begin{equation}
%\Gamma_m = \omega_m \cdot \frac{\pi^3\hbar}{120\rho c_{s,l}^5} \cdot \gamma_l^2 \left(\frac{k_B T}{\hbar} \right)^4
%\end{equation}
%where $\rho$ is the the density of GaAs, $c_{s,l}$ the sound velocity of a longitudinal wave, and $\gamma_l = (3c_{11} + c_{111}) / c_{11}$ the longitudinal Grüneisen parameter, with $c_{11}$ and $c_{111}$ the second order and third order elastic constants.

\begin{equation}
\label{eqAkhiezer}
\Gamma_m = \omega_m \cdot \frac{C_p T (\Delta\gamma)^2}{\rho \bar{c}^2} \cdot  \frac{\omega_m \tau_{ph}}{1 + \left(\omega_m \tau_{ph}\right)^2}
\end{equation}
where $C_p$ and $\rho$ are the volume specific heat and density, $(\Delta\gamma)^2$ is the variance of the Grüneisen parameter over thermal phonons involved in the process, ${3}/{\bar{c}^3} = {1}/{c_{l}^3} + {2}/{c_{t}^3}$ is the mean Debye sound velocity with $c_{l}$ ($c_{t}$) the longitudinal (transverse) velocity. The relaxation of thermal phonons occurs both in the bulk and at the resonator's surface $\tau_{ph}^{-1} =\tau_{bulk}^{-1} +\tau_{surf}^{-1}$, where $\tau_{bulk}={3\kappa}/{C_p \bar{c}^2}$ is a temperature dependent relaxation time \cite{ClelandBook} with $\kappa$ the bulk thermal conductivity, and $\tau_{surf}$ a surface relaxation time governed by the resonator geometry. In the spirit of prior works on micro and nanoscale resonators \cite{RoukesGaAs2002, kunal2011akhiezer}, we adopt the relation $\tau_{surf}=\sqrt[3]{V_{R}}/\bar{c}$ with $V_{R}$ the resonator's volume. The temperature dependance of $\tau_{ph}$ is mainly set by $\kappa$ and ${C_p}$ \cite{Blakemore1982}, and in second order by $\bar{c}$  \cite{Cottam1973}, leading to the evolution of $ \omega_{m}\tau_{ph}$ shown in inset of Fig. \ref{measurementsTED}b. The Akhiezer mechanism requires a finite variance of the Grüneisen parameter $\Delta\gamma \neq 0$,  which is approximated \cite{Saunders1974} by $(\Delta\gamma)^2=1.5 \bar{\gamma}^2$, with $\bar{\gamma}$ the average Grüneisen parameter and the factor $1.5$ taken to reproduce bulk acoustic attenuation around 300 MHz \cite{Saunders1974}. The Akhiezer prediction of Eq.\ref{eqAkhiezer} is reported in Fig. \ref{measurementsTED}b, and accounts for a first part of the phonon-phonon damping. The strain field of the RBM being non-uniform, the anharmonicity of the lattice ($ \bar{\gamma} \neq 0$) additionally induces temperature gradients within the vibrating resonator, leading to irreversible heat flows and dissipation. This thermoelastic damping (TED) \cite{Zener,Perrin1981,RoukesThermoelastic,kiselev2008phonon,kunal2011akhiezer} can be simulated by FEM, resulting in the extra contribution reported in Fig. \ref{measurementsTED}b when the temperature dependence of thermal expansion is accounted for \cite{Soma1980}. The total phonon-phonon damping is finally plotted in Fig. \ref{measurementsTED}b. It shows an overall increase with temperature, yet with no peak nor plateau. Whatever the temperature, its amplitude is also smaller than in measurements, being negligible for $T<50$ K and representing a small contribution at higher temperatures. Our models hence indicate that phonon-phonon mechanisms do not govern the dissipation of our nanomechanical resonators. This conclusion is supported by the clear difference in dissipation amplitude between the two wafers shown in Fig. \ref{measurementsTED}a, which points towards material-related effects that need to be elucidated.

%a mechanical quality factor $Q_m > 10^5$ at $300$ K and $Q_m \gg 5\cdot10^5$ for $T<100$ K.

\begin{figure*}[htpb]
%  \begin{minipage}[c]{0.7\textwidth}
    \includegraphics[width=13.3cm]{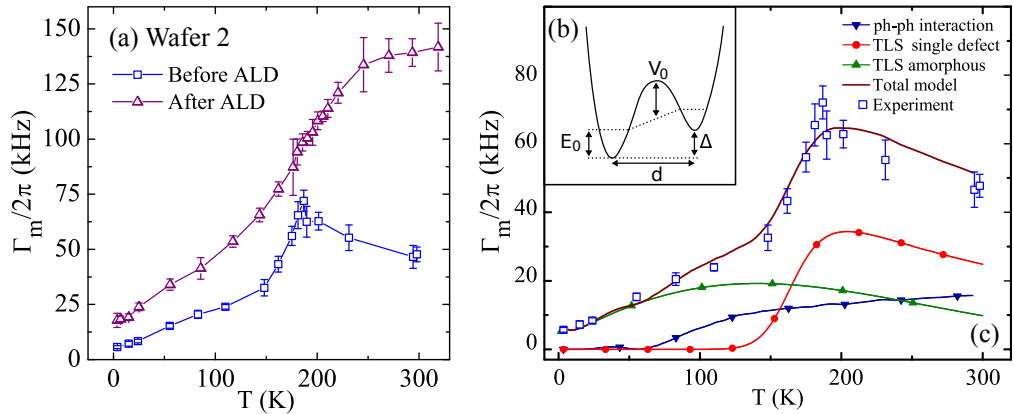}
%  \end{minipage}\hfill
%  \begin{minipage}[c]{0.25\textwidth}
\caption{(a) Intrinsic mechanical dissipation before and after ALD of $6.5$ nm of Al$_2$O$_3$. (b) Double-well model for TLS. (c) Modeling of the intrinsic mechanical dissipation in wafer 2 using phonon-phonon interactions and TLS models. \label{measurementsexp}}
%  \end{minipage}
\end{figure*}

In order to investigate the contribution of surfaces, we deposit a $6.5$ nm layer of alumina by ALD onto resonators made out of wafer 2, and compare in Fig. \ref{measurementsexp}a the temperature dependence of dissipation before and after ALD treatment. The ALD treatment increases the dissipation at all temperatures: the peak around $180$ K vanishes and the plateau-like behavior is replaced by a monotonous increase. The outcome of this trial is that surfaces play an important role in the mechanical dissipation of GaAs nano-resonators. This will be further illustrated in the analysis below.

Indeed the exact temperature dependance of $\Gamma_m$, together with its variation with the employed surface treatment or wafer, can help identifying the microscopic origin of dissipation. With this mindset, we systematically analyse our results with TLS models initially developed for amorphous materials \cite{phillips1987two,Tielburger1992,rau1995acoustic,vacher2005anharmonic}. These models depict microscopic defects and configurations as potential energy double-wells (Fig. \ref{measurementsexp}b) with the following parameters: the asymmetry $\Delta$, the barrier height $V_0$, or the well-to-well tunneling amplitude $\Delta_0 \approx {2E_0}/{\pi} \cdot \exp\left[-d\left({mV_0}/{2\hbar^2}\right)^{1/2}\right]$, with $d$ the separation between wells and $E_0$ the ground state energy of a single well \cite{phillips1987two}. For a TLS distribution $P(\Delta,\Delta_0)$, the dissipation rate is in general given by \cite{phillips1987two}:
\begin{widetext}
\begin{equation}
\Gamma_m = \omega_{m} \cdot \frac{2\eta^2N_{TLS}}{\rho c_s^2 k_BT} \iint_{\mathbb{R+}} \,d\Delta\,d\Delta_0 \frac{\Delta^2}{E^2} \sech^2\left(\frac{E}{2k_BT}\right) \frac{\omega_{m} \tau}{1+(\omega_{m} \tau)^2} P(\Delta,\Delta_0)
\label{eqphillips}
\end{equation}
\end{widetext}
with $\eta$ the deformation potential coupling of a TLS to the mechanical phonon (in eV), $N_{TLS}$ the density of TLS, $\rho$ and $c_s$ the density and sound velocity of the material, $E = \sqrt{E_0^2 + \Delta^2}$ and $\tau$ the TLS relaxation time. For $T \geq 10$ K, several energy levels of the wells are typically populated. In this so-called \emph{thermally activated regime}, the relaxation time is given by a quasi-Arrhenius law $\tau = \tau_0 e^{\frac{V_0}{k_BT}}\sech\left(\frac{\Delta}{2k_BT}\right)$ where $\tau_0^{-1}$ is of the order of the Debye frequency of the material \cite{rau1995acoustic}, and $\Delta \approx E$, leading a simplified expression of the dissipation \cite{vacher2005anharmonic}, where a distribution $P(\Delta,V_0)$ is used equivalently.
%\begin{equation}
%\Gamma_m = \omega_{m} \cdot \frac{\eta^2 N_{TLS}}{\rho c_s^2 k_BT} \int_{-\infty}^{\infty} \,d\Delta \int_0^\infty \,dV_0 \sech^2\left(\frac{\Delta}{2k_BT}\right) \frac{\omega_{m} \tau}{1+(\omega_{m} \tau)^2} P(\Delta,V_0) 
%\label{eqvacher}
%\end{equation}
%\end{widetext}

% the distribution functions of the two-levels systems. This means there is a set of TLS, with different asymmetries and barrier heights that contribute to the mechanical dissipation. Eq. \eqref{eqvacher} is a specific case of eq. \eqref{eqphillips} 

In order to fit the measured temperature-dependent dissipation, we consider two distinct distributions introduced in the literature \cite{hunklinger1976physical,vacher2005anharmonic} and sum up their contribution using the linearity of Eq.  \eqref{eqphillips}. The \emph{single defect distribution} (1) $P(\Delta^{'},\Delta_0^{'}) = \delta(\Delta^{'} - \Delta) \cdot \delta(\Delta_0^{'} - \Delta_0)$ considers both $\Delta$ and $\Delta_0$, and hence $E$, as fixed, with $\delta$ the Dirac function. This distribution assumes a single type of TLS, and makes use of an alternative deformation potential $D = \frac{\Delta}{E}\eta$ for conciseness. The \emph{amorphous distribution} (2) $P(\Delta,V_0) = f(\Delta) \cdot g(V_0)$ where $f$ and $g$ are Gaussian functions in $\Delta$ and $V_0$, with mean value $0$ and width $\Delta_1$ and $V_1$ \cite{vacher2005anharmonic}. This model is typically used for amorphous materials where a broad distribution of TLS couple to acoustic waves. In Fig. \ref{measurementsexp}c, these two contributions sum-up with the prior phonon-phonon contribution to reproduce the mechanical dissipation measured on resonators of wafer 2 in the range $T \geq 10$ K, with TLS parameters shown in Table \ref{fittingparameters}. We note that the dissipation below $10$ K is difficult to model using the thermally activated regime, such that the two lowest temperature points are fitted with a coherent version of the relaxation time \cite{phillips1987two,phillips1988} $\tau^{-1} = \frac{\eta^2\Delta_0^2E}{2\pi\rho c_s^5\hbar^4} \coth{\left(\frac{E}{2k_BT}\right)}$. The obtained level of agreement sheds light on the microscopic nature of dissipation. The mechanical damping up to 100 K is well explained by an amorphous TLS model, which suggests a role of the surface reconstruction layer, whose amorphous nature was observed by transmission electron microscopy \cite{Parrain2015}. In order to model the dissipation at higher T, the single defect model must however be used on top. The related defect has an activation energy of $\approx 0.1$ eV, consistent for example with GaAs $DX$ centers \cite{yamanaka1987electron}. For a deformation potential $\approx 10$ eV, the inferred density of TLS is of $\approx 10^{19}$cm$^{-3}$, well above the unintentional doping of our epitaxial GaAs ($10^{14}$cm$^{-3}$ range), suggesting the formation of TLS at the surfaces in a density superior to the bulk. The localization of TLS can further be investigated by looking at ALD-treated resonators and resonators fabricated out of wafer 1 (see Table \ref{fittingparameters}). The ALD surface treatment appears to modify both the amorphous and single defect distribution, indicating again that both types of TLS can be at surfaces. It enhances the density of amorphous TLS, which is consistent with the amorphous nature of deposited alumina; but decreases the density of single defect TLS, which is reminiscent of the passivation action of ALD \cite{guha2017surface}. The superior mechanical dissipation in wafer 1 compared to 2 seems to originate from a different density of amorphous TLS, which is again consistent with their localization at surfaces, since surface absorption of photonic resonators fabricated out of wafer 1 was also superior to that of wafer 2 \cite{guha2017surface}.	

%$g(V_0) \propto \frac{1}{V_0}\left(\frac{V_0}{V_1}\right)^{-1/4} \exp\left(-\frac{V_0^2}{2V_1^2}\right)$. 
%\begin{enumerate}
%\item \emph{Single defect distribution} 
%\begin{equation}
%P(\Delta^{'},\Delta_0^{'}) = \delta(\Delta^{'} - \Delta) \cdot \delta(\Delta_0^{'} - \Delta_0)
%\end{equation}

%\item \emph{Amorphous distribution} 
%\begin{equation}
%P(\Delta,V_0) = f(\Delta) \cdot g(V_0)
%\end{equation}
%where $f$ and $g$ are Gaussian functions in $\Delta$ and $V_0$, with mean value $0$ and width $\Delta_1$ and $V_1$ \cite{vacher2005anharmonic}. $g(V_0) \propto \frac{1}{V_0}\left(\frac{V_0}{V_1}\right)^{-1/4} \exp\left(-\frac{V_0^2}{2V_1^2}\right)$. This model is typically used for amorphous materials where a broad distribution of TLS couple to acoustic waves.
%\end{enumerate}

\begin{table*}
\caption{Fitting parameters for TLS models in the thermally activated regime.\label{fittingparameters} }
\begin{tabular}{l || c | c ||c | c || c | c}
\hline
& \multicolumn{2}{c||}{Wafer 2 (without ALD)} & \multicolumn{2}{c||}{Wafer 2 (with ALD)} & \multicolumn{2}{c}{Wafer 1 (without ALD)} \\ \cline{2-7}
& Amorphous & Single defect & Amorphous & Single defect & Amorphous & Single defect \\ \hline
$\Delta_1$ (J) & $1000\cdot k_B$ & - & $1000\cdot k_B$ & - & $1000\cdot k_B$ & - \\
$V_1$ (J) & $1500\cdot k_B$ & - & $1500\cdot k_B$ & - & $1500\cdot k_B$ & - \\
$\tau_0$ (s) & $4\cdot 10^{-13}$ & $4.25\cdot 10^{-12}$ & $4\cdot 10^{-13}$ & $4.21\cdot 10^{-10}$ & $4\cdot 10^{-13}$ & $4.25\cdot 10^{-12}$ \\
$E$ (eV) & - & $0.1$ & - & $0.05$ & - & $0.1$ \\
$\eta^2N_{TLS}$ (eV$^2 \cdot$m$^{-3}$) & $1.82 \cdot 10^{25}$ & - & $1.11 \cdot 10^{26}$ & - & $9.59 \cdot 10^{25}$ & - \\
$D^2N_{TLS}$ (eV$^2 \cdot$m$^{-3}$) & - & $9 \cdot 10^{26}$ & - & $6.48 \cdot 10^{25}$ & - & $1.98 \cdot 10^{27}$ \\
\hline
\end{tabular}
\end{table*}

%\begin{enumerate}
%	\item At low temperature, typically $T \leq 10$ K, there is \emph{coherent tunneling} between the two wells, and the double-well can be described to a good agreement with the same formalism as a $1/2$-spin. Hence, it has a relaxation time:
%\end{enumerate}

% \begin{table}
 %\caption{Fitting parameters for the Two-Level System models.\label{fittingparameters} }
 %\begin{tabular*}{\columnwidth}{@{\extracolsep{\fill} } l | c | c | c }
%\hline\hline
%& Coherent  & Thermal  & Single defect\\
%\hline
%$\gamma$ (eV) & $5$ & $1.35$ & -\\
%$N_{TLS}$ (m$^{-3}$) & $\approx 7\cdot10^{21}$ & $\approx 1\cdot10^{25}$ & $\approx 1\cdot10^{24}$\\
%$\Delta_0$ (J) & $3\cdot k_B$ & - &-\\
%$\Delta_1$ (J) &$10\cdot k_B$ & $1000\cdot k_B$ &-\\
%$V_0$ (J)&- & $1500\cdot k_B$ & -\\
%$\tau_0$ (s) & - & $4\cdot 10^{-13}$ & $4.25 \cdot 10^{-12}$\\
%$E$ (eV)& - & - & $0.1$\\
%$D$ (eV)& - & - & $30$\\
%$N_{TLS} \cdot \gamma^2$ (eV$^2 \cdot$ m$^{-3}$)  & $1.75\cdot10^{23}$ & $1.82\cdot10^{25}$ & -\\
%$N_{TLS} \cdot D^2$ (eV$^2 \cdot$ m$^{-3}$) & - & - & $9\cdot10^{26}$\\
 %\hline\hline
 %\end{tabular*}
 %\end{table}

In summary, we have reported a systematic study of intrinsic nanomechanical dissipation in GaAs resonators. Microscopic models indicate that two-level systems dominate damping at any temperature between 3 and 300K. While in conflict with the common sense that crystalline devices are less affected by TLS than their amorphous counterparts \cite{WeigTLS2014,SchmidTLS2014}, this conclusion is consistent with the presence of an amorphous reconstruction layer at their surface. Such a layer already rules the optical dissipation of high-Q GaAs resonators with large surface to volume ratio \cite{Parrain2015}, and we bring here a series of evidences that TLS impacting their nanomechanical dissipation mainly localize at surfaces as well. Our models anticipate that the freezing of these fluctuating TLS would be beneficial, predicting a mechanical quality factor $Q_{m}$ beyond $10^{9}$ at 10 mK. The related Q-frequency product $Q_{m}\times f_{m}$ would reach the $10^{17}-10^{18}$ range for GaAs resonators, equalling the performances of other crystalline devices in quartz \cite{Tobar2012} and silicon \cite{Meenehan2015}. Ultra-low temperature experiments, possibly below the milliKelvin, may ultimately reveal how far the performances of nanomechanics can be pushed, for metrological and quantum applications.

This work was supported by the European Research Council (ERC) through the Ganoms project (No.306664). The authors thank Bernard Perrin and Eddy Collin for fruitful comments.

% If in two-column mode, this environment will change to single-column format so that long equations can be displayed. 
%\begin{widetext}
%$$\mbox{put long equation here}$$
%\end{widetext}

% Figures should be put into the text as floats. 
% Use the graphics or graphicx packages (distributed with LaTeX2e).
% Put the label that you will use with \ref{} command in the braces of the \label{} command.
%
% \begin{figure}
% \end{figure}

% \begin{table}
% \end{tabular}
% \end{table}

%\begin{acknowledgments}
% Put your acknowledgments here.
%\end{acknowledgments}

\bibliography{bibliography}

\end{document}